\documentclass{iaus}
\usepackage{graphicx,natbib}

\title[Simulating the IMF] 
{Simulations of the IMF in Clusters}

\author[Ralph E. Pudritz]   
{Ralph E. Pudritz}  

\affiliation{Origins Institute, McMaster University, ABB 241,\\
Hamilton, Ontario L8S 4M1, Canada \\ email: {\tt pudritz@mcmaster.ca} \\[\affilskip]
}

\pubyear{2010}
\volume{270}  
\pagerange{119--126}
\jname{Simulations of the IMF in Clusters}
\editors{J. Alves, B.G. Elmegreen, J. Miquel, \& V. Trimble, eds.}
\begin{document}

\maketitle

\begin{abstract}
We review computational approaches to understanding the 
origin of the Initial Mass Function (IMF) during the formation of star clusters. 
We examine the role of turbulence, gravity and accretion, equations of state,
and magnetic fields in producing the distribution of core masses - the Core Mass Function (CMF).
Observations show that the CMF is similar in form to the IMF.  We focus on feedback
processes such as stellar dynamics, radiation, and outflows
can reduce the accreted mass to give rise to the IMF.   Numerical 
work suggests
that filamentary accretion may play a key role in the origin of the IMF.

\end{abstract}

\firstsection 
\section{Introduction}

The Initial Mass Function (IMF)  
plays a central role in astrophysics because it  
encapsulates the complex physics of star formation. 
Observations  
suggest that the IMF can be variously described as a piece-wise power-law 
\citep{Kroupa:2002}, a lognormal \citep{Miller/Scalo:1979}, 
or a lognormal distribution with a high-mass power-law tail \citep{Chabrier:2003}.  
The high mass behaviour of the IMF \cite{Salpeter:1953}   
is a power-law,
$dN \propto m^{-2.3} dm$ for stellar masses $m \ge 0.5 M_{odot}$.  
The peak mass 
for isolated stars in the galactic disk is $ 0.1 M_{\odot}$ 
and $ 0.2-0.3 M_{\odot}$  \citep{Chabrier:2003} for the bulge. 
The form of the IMF is similar in many different galactic and
extragalactic environments such as globular clusters, wherein
one has a large range of metallicities and concentrations 
\citep{Paresce/DeMarchi:2000}. 
The current evidence therefore tends to support the
notion that the IMF is universal.

What physical processes produce the IMF?
Observations show that it emerges during the early stages 
of the formation of star clusters
\citep{Meyer/etal:2000,Zinnecker/etal:1993}. 
Young stars are formed within gravitationally bound subunits of 
a cluster-forming environment known as "cores" whose mass distribution - 
the core mass function (CMF) - strongly resembles the IMF.
Cores are closely 
associated with filaments, as recent 
observations using the Spitzer and Herschel observatories clearly show 
\citep{Andre/etal:2010}.  
Competing physical processes such as turbulence, gravity, cooling and thermodynamics, 
as well as magnetic fields play significant roles in 
building the CMF, as well as filaments, within molecular clouds.
Feedback processes such 
as radiation from massive stars, jets and outflows, as well as stellar 
dynamics serve to truncate the accretion of material onto stars and their
natal disks.  These may lead to the emergence of the IMF from the CMF.  

This review focuses on the critical role that computation is playing  
in exploring the origin of the IMF in clusters.  We focus first on 
the processes leading to the CMF, and then discuss feedback processes 
that may convert the CMF into the IMF.  
Recent reviews of the computational aspects of the IMF may be found
in \citet{MacLow/Klessen:2004},  
\citet{Bonnell/etal:2007}, 
\citet{Larson:2007}, 
and \citet{Klessen/etal:2009}.
The theory of the IMF is covered by Hennebelle (this volume), 
and \citet{McKee/Ostriker:2007}.  

\section{Structure formation and the IMF} 

Structure formation in the diffuse ISM as well as the dense
molecular medium is a shocking affair.  
Diffuse atomic hydrogen near the midplane of our Milky Way 
is observed to be organized as a plethora of filamentary structure, 
bubbles, supernova remnants, HII regions \citep{Taylor/etal:2003}.  
Filamentary structure also characterizes giant molecular clouds -
as seen in the large scale extinction maps of the Orion 
and Monocerus clouds \citep{Cambresy:1999}. 
On smaller scales, a 850 micron continuum map of 
a 10pc region in Orion shows that Bonner-Ebert like cores
are associated with filaments \citep{Johnstone/Bally:2006}.  
Herschel observations, 
such as those of the IRDC filament known as the "snake", show 
massive stars and a star cluster in formation \citep{Henning/etal:2010}.  

How can cloud structure be characterized?  
One fruitful approach 
is to measure the probability distribution functions (PDFs) 
of the cloud column density of all of the gas in 
a given molecular cloud 
which is readily measured directly from the extinction data. 
This PDF for clouds without star formation (such as Lupus V and the 
Coal Sack clouds) turns out to be lognormal, 
whereas that
of star forming clouds is a lognormal plus high mass, power-law tail 
\citep{Kainulainen/etal:2009}.  This is interpreted as arising from the 
effects of gravity which drives collapse.  

Filaments have various origins and arise in supersonic turbulent
media due to the intersection of shocks waves, by gravitational break-up of self-gravitating
sheets, or by thermal instabilities of various kinds.  
A physically plausible picture for the origin of the CMF is that  
filaments with sufficiently high values
of their mass per unit length produce cores by gravitational instability.  
Gravitational fragmentation of 
filaments is well studied theoretically \citep{Nagasawa:1987, Fiege/Pudritz:2000} but
has received renewed emphasis with the Herschel observations.

The similarity in the functional form of the CMF and the IMF has been observed in many clouds 
starting with the study of \citet{Motte/etal:1998} in $\rho$ Oph.  
The CMF in the Pipe Nebula, as another example,
can be shifted into the IMF
by converting a fraction $\epsilon \simeq 1/3 $ of its mass 
to stars \citep{Alves/etal:2007}.

\section{From clouds to the CMF}

\subsection{Turbulence}

Supersonic turbulence rapidly compresses gas into a hierarchy of sheets and filaments 
wherein the denser gas undergoes gravitational collapse to form stars  
\citep{Porter/etal:1994,
Vazquez-Semadeni/etal:1995, 
Ostriker/etal:1999, 
Klessen/Burkert:2001, 
Padoan/etal:2001,
Bonnell/etal:2003, 
Tilley/Pudritz:2004, 
Krumholz/etal:2007}. 
Supersonic gas is also highly dissipative and without constant replenishment,
damps within a crossing time. 
There are many sources of turbulent motions that can affect molecular clouds, 
such as galactic
spiral shocks in which most giant molecular clouds form, 
supernovae, expanding HII regions, radiation pressure, cosmic
ray streaming, Kelvin-Helmholtz and Rayleigh-Taylor instabilities, 
gravitational instabilities, and 
bipolar outflows from regions of star formation \citep{Elmegreen/Scalo:2004}.

Supernova driven bubbles and turbulence in the galactic disk have been simulated 
by several groups.  The work of \citet{Avillez/Breitschwerdt:2004} simulates the 
high resolution (down t0 1.5 pc scales) global structure
of the ISM as a function of the supernova rate.   
Densities range over six orders of magnitude, 
$10^{-4} \le n \le 10^2$ cm$^{-3}$ and multi-phase density PDFs are given.  
Simulations of the global structure of a supernova lashed, 
multi-component, instellar medium 
\citep{Tasker/Bryan:2008,  
Wada/Norman:2007} find that the density PDF follows a lognormal distribution.

Supersonic turbulence 
produces  hierarchical structure that can be described by a lognormal
distribution \citep{Vazquez-Semadeni:1994}.  A lognormal arises 
whenever the probability
density of each new step in density increment in the turbulence is independent of the
previous one.  As an example, consider a medium that undergoes a series
of random shocks whose strengths are uniformly distributed 
\citep{Kevlahan/Pudritz:2009}.    
The density at any point is the product of the 
shock-induced, density jumps.  Taking the log of this relation, and then the limit of 
a large number of shocks, the central limit theorem then shows 
that the log of the gas density should be normally distributed 
- hence the lognormal.  It can also be shown mathematically
that the convergence of the distribution for a finite number of shocks is very rapid - 
just 3 or 4 shocks will give a distribution that is very highly converged to a lognormal. 
  
Lognormal behaviour for the CMF 
has been found in a wide variety of simulations using various types of codes (eg. SPH, AMR) and 
setups (driven or not driven, periodic boxes, initial uniform spheres, etc.). 
Early results showed that lognormal behaviour in periodic box simulations 
is independent of details on how the turbulence is driven 
\citep{Klessen:2001}.

It is interesting that lognormal 
distributions appear across science (eg. physics, biology, medicine, etc.), not 
just in fluid mechanics.   
The key
difference between normal and lognormal distributions in general is that
the former arises for additive processes, whereas the latter
arise in multiplicative ones \citep{Limpert/etal:2001}.  
As a concrete example, consider a simple dice game 
where one first adds the values on the faces of two thrown
dice - the distribution of results (ranging from 2 to 12) 
is a normal distribution whose
mean is 7.  If one multiplies the two values however, the resulting distribution
of numbers is highly skewed (ranging between 1 and 36) and is described
by a lognormal.

Thus, lognormals characterize  
structure in the diffuse ISM as well as in  
denser molecular gas because shocks are 
the dominant process for configuring the gas. This is independent
of exactly how the shocks are produced.  Of physical significance
are the mass of the peak of this distribution, and its width $\sigma_o$.  
The latter depends on both the thermal state of the 
gas as well as the rms Mach number of the turbulence.  
The standard deviation $\sigma_o$ is found from the 
simulations, and takes the form;
$\sigma_o^2 = ln (1 + b^2 M^2) $ where M is the rms Mach number of the
turbulence and $b \simeq 0.5$ is a fitting parameter      
\citep{Padoan/etal:1997}. 

\subsection{Gravity and accretion}

Adding gravity to turbulence changes this distribution - a power-law tail appears
at the high mass end of the simulated CMF 
\citep{Li/etal:2003, 
Tilley/Pudritz:2004}.  
In the semi-analytic treatment of \citet{Padoan/Nordlund:2002},   
the power-law arises because of the turbulence spectrum of the turbulence.    
In \citet{Hennebelle/Chabrier:2008}, a Press-Schecter formalism
is adopted to argue that a lognormal plus power-law behaviour is the consequence
of imposing a star formation threshold (eg. Jeans' criterion) and gravity
at high mass.

Much of the debate concerning the origin of the CMF and IMF has focused on whether
stars form by competitive accretion \citep{Bonnell/etal:2001}  
or by the gravitational collapse of discrete
cores \citep{Krumholz/etal:2005}.  
Cluster formation simulations often use initial top-hat density profiles (uniform
spheres) that are chosen to mimic the observed initial conditions of cluster forming clumps 
(eg. a hundred solar masses of material, at temperature of 10K, size of half a pc,
and mean density of at least $10^5$ cm$^{-3}$). Such simulated clumps 
start to undergo global gravitational collapse
in less than a free-fall time 
(eg. \cite{Bonnell/etal:2001}, 
\cite{Tilley/Pudritz:2004}) 
as the turbulent
energy is dissipated.  The collapsing background ramps
up the density of the gas including those in fluctuations such as the  
filaments.  This drives up the accretion of gas into the filaments
pushing some of them towards gravitational fragmentation.
Sink-particles (taken as proxies for cores) first appear 
within the filaments.
Cores will collapse more quickly than the collapsing
background clump because they are denser and therefore have
shorter free-fall times.  The the collapse carries the collection
of sink particles and filaments into 
the ever deepening potential well of the clump.     

Cores are not isolated objects within filaments.  Rather, they
continue to undergo considerable filamentary accretion.  
The first objects to 
appear within most simulations generally become the most massive cores.  
In \cite{Banerjee/etal:2006}, the 
rapid growth of the first, and most massive star by filamentary accretion was followed
with an AMR (FLASH) code.  These simulations showed that the filament was formed at the 
intersection of two sheets- shocks.  
   
\subsection{Equations of state}

The thermal state
of the bulk gas is a major factor in determining the local Jeans mass
from point to point in molecular clouds.  
Both theory and simulations show that
equation of state plays a very important 
role in controling the gravitational fragmentation of the gas.
The local Jeans mass scales with the local temperature and 
density as $M_J \propto T^{3/2} \rho^{1/2}$.  
For simple polytropic equations of state $P \propto \rho^{\gamma}$, the Jean's 
mass can be written purely as a power law of the density;
$M_J \propto \rho^{3/2(\gamma - (4/3))}$.  
This scaling suggests that for $\gamma > 4/3$, the Jeans mass
increases with density, which puts an end to fragmentation. 
Indeed, simulations show that
strong fragmentation prevails for $\gamma \le 1$, gets progressively 
weaker for $\gamma > 1$, and stops altogether for $\gamma > 1.4$ \citep{Li/etal:2003}.

Local energy 
sources can raise the fragmentation mass by changing the 
temperature of the region.  Thus, the accretion luminosity released by 
massive stars in particular, must reduce the degree of fragmentation
within a localized region around such a heat source,
as has been demonstrated by several groups \citep{Krumholz/etal:2007}.
For massive stars, this region is limited in extent - roughly 1000 AU or so
(see \S 4).

\subsection{Magnetic fields}

Magnetic fields play several different roles in star formation.  As has long
been known, they can control the gravitational stability of a gas if their
energy density exceeds that of gravity.  This is formalized by 
mass to flux ratio; $\Gamma = 2 \pi \sqrt{G} \Sigma / B = 1.4 \beta^{1/2} n_J^{1/3}$ 
where $n_J$ is the number of Jeans masses and $\beta$ is the ratio of gas to magnetic
pressure.  The fragmentation of a uniform cloud is highly suppressed for 
subcritical clouds ($\Gamma < 1$).  Simulations of slightly supercritical clumps
in uniform magnetic fields show that the field channels the collapse into large
sheets that are perpendicular to the direction of the field.  Slightly more
supercritical clouds however, break up into more substructure including filaments
\citep{Tilley/Pudritz:2007}.  
Turbulence creates a very broad range of magnetizations of cores.  This is because 
shocks
sweep material along field lines where it accumulates in filaments - increasing
the mass to flux ratio in those regions, and greatly reducing the mass to 
flux ratio in the more diffuse zones left behind.  Simulations   
show that initially supercritical
magnetized clouds results in cores that range from critical to strongly 
supercritical \citep{Padoan/Nordlund:2002, Tilley/Pudritz:2007}, in 
agreement with the observations 
\citep{Crutcher:2007}.  
This may also account for the fact that magnetic fields
are more dominant in the diffuse gas than in molecular gas. 

When turbulence is
added to a subcricital cloud, the column density PDF of a cloud
is a lognormal but with a very small standard deviation.  
When ambipolar diffusion is added, the lognormal broadens considerably 
\citep{Nakamura/Li:2008}).  
Supercritical subregions can form within the subcritical cloud, and it is within
these regions that star formation can proceed, albeit at a very heavily reduced rate.

The second major aspect of magnetic fields is that twisted fields exert torques on gas 
and can therefore transport angular momentum away from spinning bodies.  
The origin of angular momentum in supersonic turbulence is that oblique shocks 
produce spinning cores \citep{Jappsen/etal:2004, Tilley/Pudritz:2007}).    
Turbulence simulations produce 
a broad distribution of angular momenta of cores ranging over nearly 
two orders of magnitude.
This implies that a broad distribution of disk sizes should result from the 
collapse of these systems - from very small to very large disks.  The distribution
of angular momentum vectors is also quite varied and is not particularly aligned
with the filament principal axes.  

Magnetic fields participate in several kinds of braking.  On the level of the cores,
torsional Alfv\'en waves have long been known to be able to extract significant amounts
of angular momentum \cite{Basu/Mouschovias:1994}.  The collapse of 
rotating cores produces magnetized outflows as demonstrated in 
a variety of initial core models; cylinders \citep{Tomisaka:2002},
Bonner-Ebert
spheres \citep{Banerjee/Pudritz:2006}, uniform spheres \citep{Hennebelle/Fromang:2008}, 
and singular isothermal
spheres \citep{Mellon/Li:2008}.  Early outflows could sweep up
significant amounts of material and have been implicated as the basic physics in 
the CMF to IMF efficiency factor $\epsilon$ (see  
\S 4.3).  

Finally, the combination of radiative and MHD effects (RMHD) 
limits the fragmentation of cluster-forming gas.  
Attempts to model RMHD on cluster scales have been made 
by \cite{Price/Bate:2009} who used Euler potentials to approximate the MHD in SPH.  This
work shows that the MHD in their code strongly supresses fragmentation even for clouds that
are fairly supercritical (eg. $\Gamma = 3$). On smaller scales, grid-based RMHD methods applied to 
collapse and outflows show that radiative heating (in flux limited diffusion
limit) makes substantial changes to the extent of outflows \citep{Commercon/etal:2010, Tomida/etal:2010}.      

\section{Feedback: from the CMF to the IMF}

\subsection{Stellar dynamics and filamentary accretion}

\begin{figure}
\begin{tabular}{cc}
\includegraphics[width=6cm]{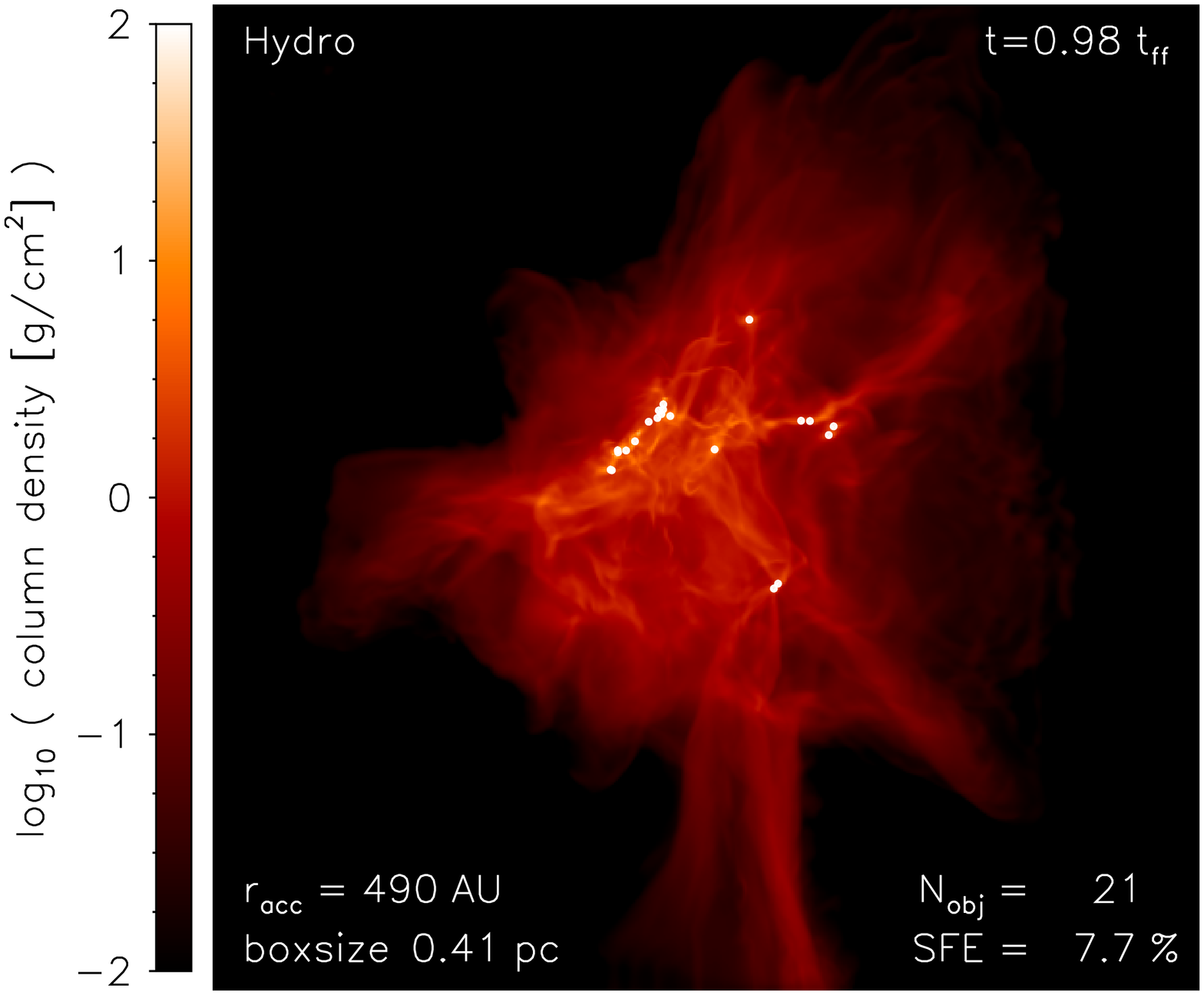} &
\includegraphics[width=6cm]{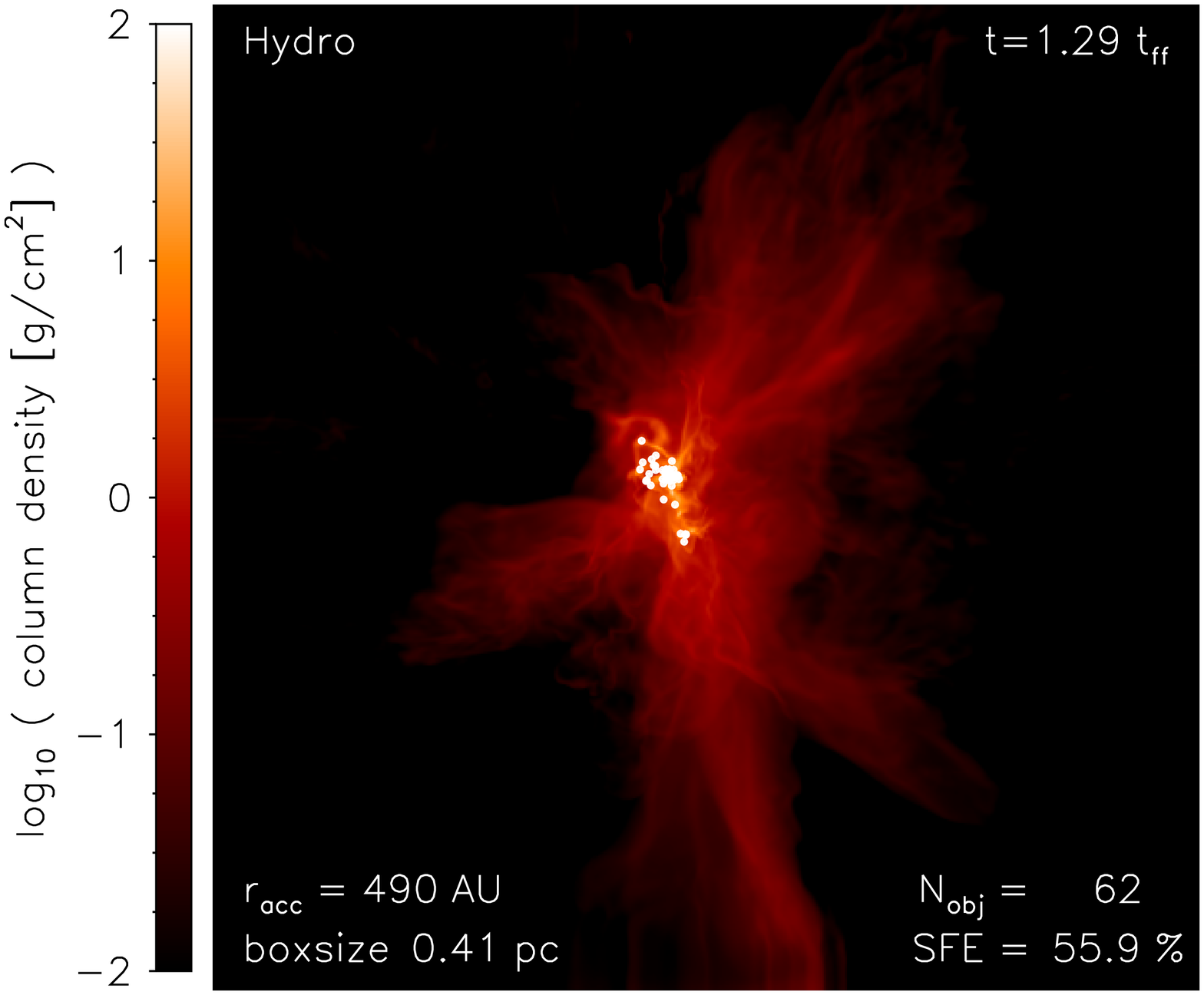} \\
Figure 1a & Figure 1b \\
\includegraphics[width=6cm]{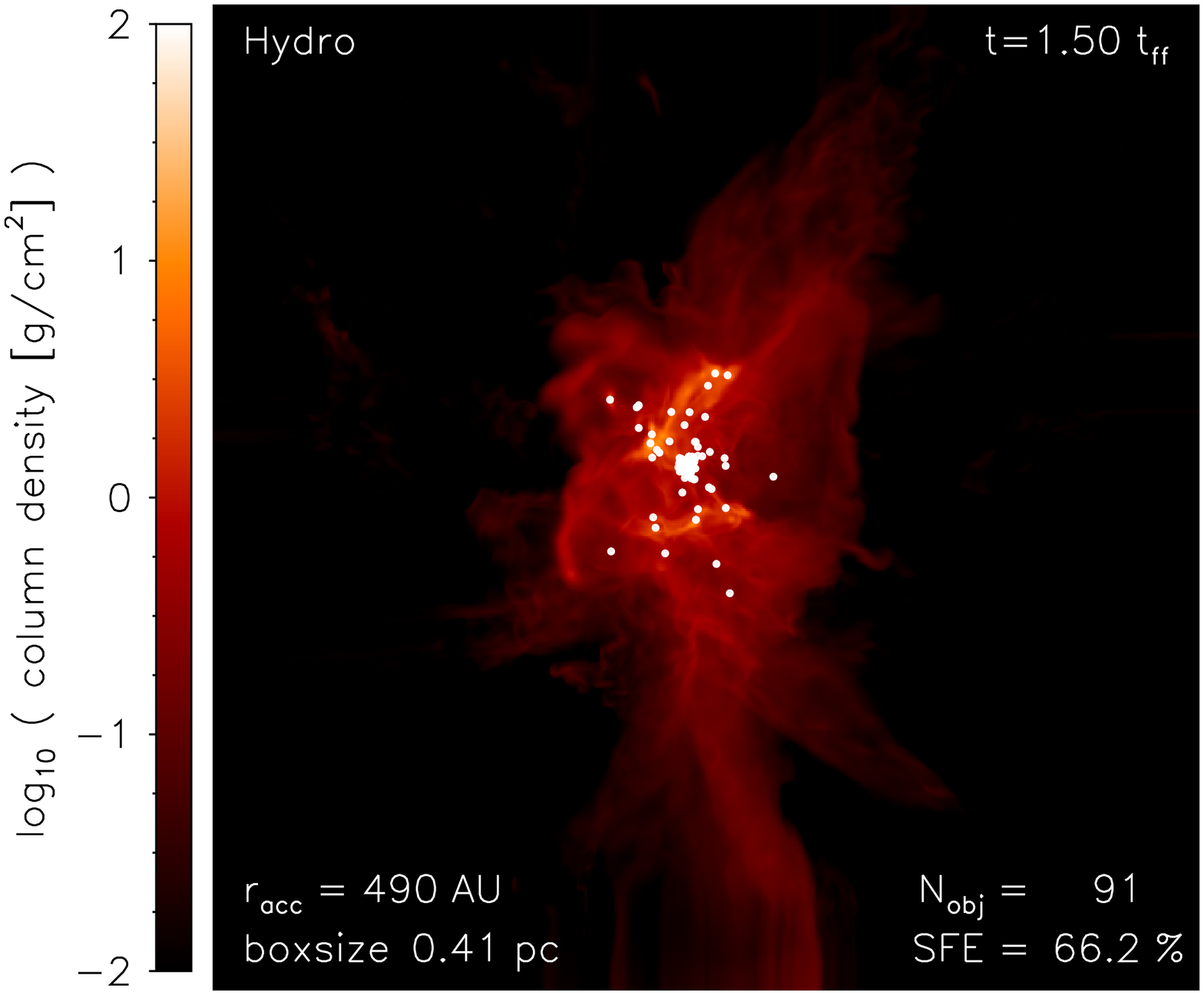} &
\includegraphics[width=6cm]{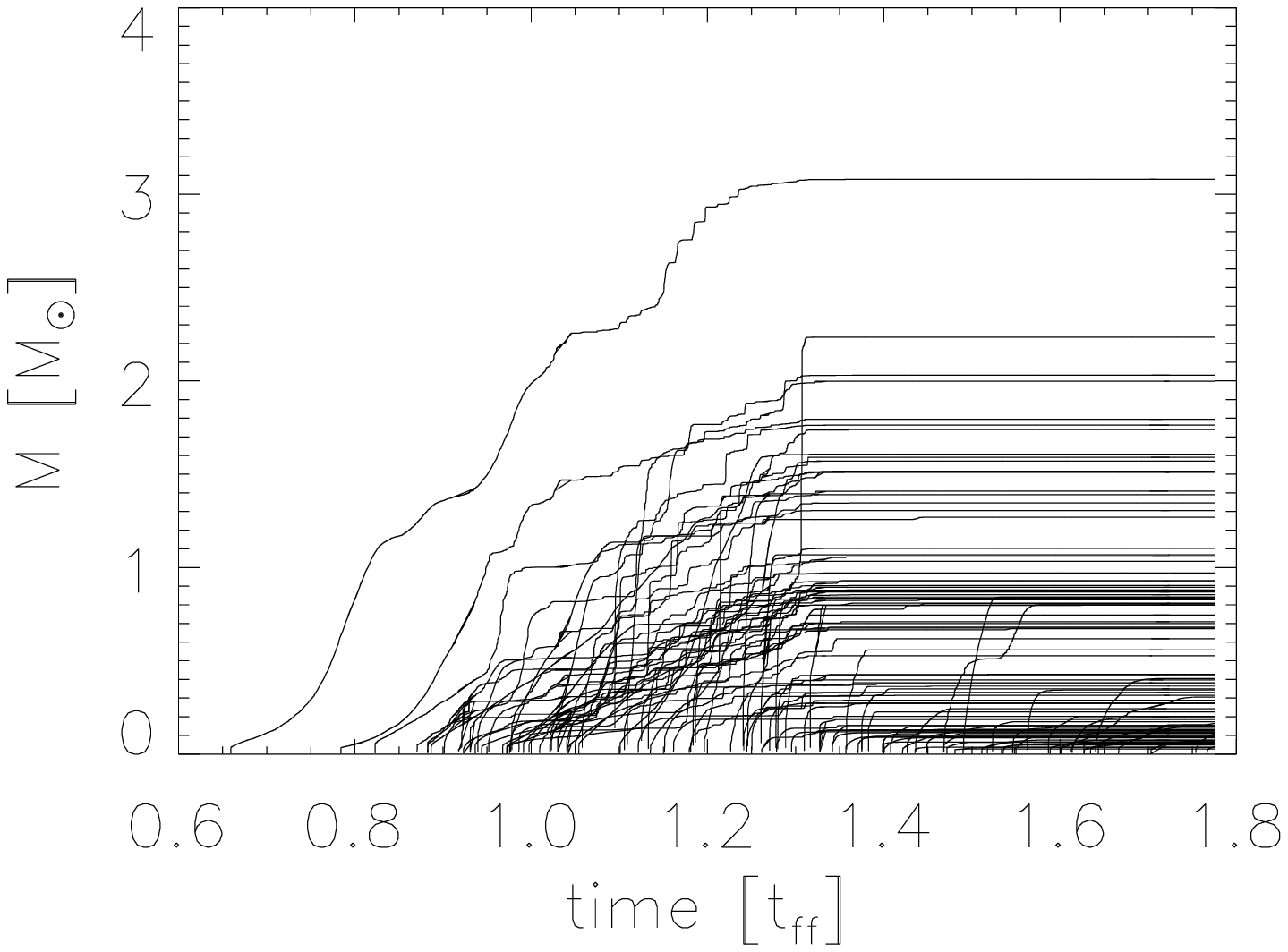} \\
Figure 1c & Figure 1d 
\end{tabular}
\caption{\label{fig:1}Cluster formation and filamentary accretion.  Top
left: sink particles form in filaments.  Top right: sink particles
begin dynamical interaction.   
Bottom left: cluster dynamics, accretion has ceased. Bottom right:
mass accretion history for each sink particle - accretion ends when
dynamical interactions begin \citep{Duffin/etal:2010}}.   
\end{figure}

The collapse of $10^2 M_{\odot}$ clumps 
pulls both the material in the filaments
and the sink particles deeper into the central potential well.  The sink-sink interactions,
which are modeled to be N-body gravity then come into play and are responsible for the
creation of a small stellar cluster.  The competitive accretion scenario predicts that these
objects compete for gas in the dense centre of the gravitational potential well, 
resulting in an IMF \citep{Bonnell/etal:2001}.  
For larger clumps ($10^3 - 10^4 M_{\odot}$) the evolution of a cluster forming region  
is a hierarchical process in which subclusters form and 
ultimately merge \citep{Bonnell/etal:2003}. The resulting 
stellar interactions are frequent and close enough to truncate
protostellar disks.  

Sink particles as implemented by \citet{Federrath/etal:2010} 
have the additional desirable feature that they form in local potential minima  
giving them a more hydrodynamic character.  This has 
an important consequence.   
As opposed to competitive accretion onto particles in a general
potential well, filamentary 
accretion largely ends when the N-body interactions become strong enough to 
kick the sink particles out of their feeding filaments
Evidence for filamentary accretion is seen 
in the simulations of \cite{Duffin/etal:2010}, and shown in  
Figure 1 (featuring the collapse of a $100 M_{\odot} \simeq 100 M_J$  
initial tophat clump). The first frame, 
shows that sink particles form in filaments.  The second shows that dynamical interactions
have started between them as the collapse of the clump proceeds, and the third shows
the dynamical
end state.  The final frame shows that the accretion histories of each particle shuts off
when when dynamical interactions become important.

\subsection{Radiative feedback} 

Comprehensive 2D simulations of massive star formation 
including full frequency, radiative feedback effects  
and dust are presented by 
\citet{Yorke/Sonnhalter:2002}.  
For cluster formation, a 3D treatment of radiative transfer in a highly inhomogenous
medium is essential.  Since \citet{Krumholz/etal:2007}, it has become widely appreciated
that radiative heating of the gas is needed to prevent excessive gravitational fragmentation.  
Radiative    
heating of the gas
prevents the filaments from fragmenting
as much as they might.  Their gas drains primarily into the central massive forming 
disk and star, and fragmentation out to 1000 AU scales is prevented.  These calculations still
invoke grey, flux limited diffusion so there is still a need to examine the role of 
different frequency regimes in this process.  
Simulations by \cite{Bate:2009} confirm this picture using SPH techniques - wherein
suppression of brown dwarfs by a factor of 4 is often observed compared to simulations
without radiative feedback.  
The ionizing radiation from massive young stars becomes important 
during cluster formation as HII regions drive hot ioinized flows in the cluster environment.
Results show that companions to massive stars limit the growth of the latter by a 
starvation effect 
\citep{Peters/etal:2010}.    

\subsection{Outflows}

Outflows carry substantial amounts of mechanical energy which, 
if coupled into clump dynamics, could affect the IMF.
Two aspects of feedback are important outflows in a cluster region
\citep{Klessen/etal:2009}.
The first is at the core level, in which an outflow
removes the collapsing gas at some efficiency.  If this
efficiency is quite high (so that $\epsilon \simeq 1/3 $), then one may 
resolve the hypothesized conversion of the CMF to IMF \citep{Matzner/McKee:2000}. 
Simulations show that the efficiency may be much lower \citep{Duffin/etal:2010b}. The 
second is at the level of the clump and the question as to whether or not 
the collection of outflows 
can continue to excite turbulence and thereby regulate 
cluster formation \citet{Norman/Silk:1980}.  

The situation at the local level is still unclear.  Simulations by \citet{Banerjee/etal:2007}
show that supersonic turbulence is not driven by magnetized jets.  
At the level of clumps, \cite{Nakamura/Li:2008} included ambipolar diffusion
and outflows in a cloud that is nearly critical ($\Gamma = 1.1$ initially), and found, not 
unexpectedly, that the magnetic fields regulated a rather slow rate of star formation.  
On the other hand, \citet{Wang/etal:2010} used a mass to flux of $\Gamma = 1.4$ in a
code without AD, and observed long time regulation of the cluster formation by outflows
which maintained turbulence.  These results may all depend upon the limited modeling of the 
the full dynamics of outflows that have been incorporated into the simulations.    

\section{Synthesis: the IMF of clusters}

Numerical simulations have become the primary tool 
with which to investigate the origin of the IMF in clusters. 
Turbulence, the equation of state, and gravity play the key roles of filamenting
the gas and breaking it into accreting cores describable by a lognormal distribution
with high mass power-law tail.  Radiative feedback controls fragmentation rates
at the low mass end of the IMF whereas feedback by outflows may not be
as efficient as previously claimed in converting the CMF to the IMF.  
Filamentary structure from clouds to clumps may turn out 
to play a key role in the entire process.


\bibliographystyle{apj}
\bibliography{bib_imf}


\end{document}